# A comparison of citation-based clustering and topic modeling for science mapping

Qianqian Xie, Ludo Waltman
*q.xie@cwts.leidenuniv.nl; waltmanlr@cwts.leidenuniv.nl*
Centre for Science and Technology Studies (CWTS), Leiden University, Leiden, The Netherlands

## Abstract

Understanding the different ways in which different science mapping approaches capture the structure of scientific fields is critical. This paper presents a comparative analysis of two commonly used approaches, topic modeling (TM) and citation-based clustering (CC), to assess their respective strengths, weaknesses, and the characteristics of their results. We compare the two approaches using cluster-to-topic and topic-to-cluster mappings based on science maps of cardiovascular research generated by TM and CC. Our findings reveal that relations between topics and clusters are generally weak, with limited overlap between topics and clusters. Only in a few exceptional cases do more than one-third of the documents in a topic belong to the same cluster, or vice versa. For TM the presence of highly similar topics is a considerable challenge. A strength of TM is its ability to represent societal needs related to cardiovascular disease, potentially offering valuable insights for policymakers. In contrast, CC excels in depicting the intellectual structure of cardiovascular diseases, with a strong capability to reflect scientific micro-communities. This study deepens the understanding of the use of TM and CC for science mapping, providing insights for users on how to apply these approaches based on their needs.

**Keywords** Topic modeling, Citation-based clustering, Science mapping, Cardiovascular research

## 1 Introduction

Science maps are visual representations of the intellectual structures, cognitive structures, social structures, and dynamics of science (Petrovich, 2020; Gläser et al., 2017). They demonstrate how disciplines, fields, subject areas, authors, keywords, or publications are related (Börner, Chen and Boyack, 2005; Osinska and Malak, 2016; Rafols, Porter and Leydesdorff, 2010; Small, 1999). Consequently, science maps are useful tools in the sociology of science and science policy (Petrovich, 2020) and have drawn considerable interest in recent decades.

Bibliometricians and scientometricians have developed various mapping methods and techniques to generate science maps. These mapping approaches employ various data models, language processing algorithms, network analyses, and visualization techniques (Gläser et al., 2017). Science maps generated using different mapping approaches may capture the intellectual structure of science at different levels and from different perspectives, even when using the same data. This may have a significant influence on decision-making processes in which researchers, science managers, analysts or policymakers use science maps as a basis for their decisions.

Consequently, it is pivotal for users to understand how solutions produced by different approaches differ from one another. Which aspects of the results are robust or specific? And how to choose between different approaches and interpret the results effectively (Gläser et al., 2017; Velden et al., 2017a)? These considerations are crucial in determining what role science maps can play to support decision making. By selecting and applying appropriate methods, the accuracy and reliability of science maps can be improved, thereby better supporting decision-making processes.

Researchers still have only a limited understanding of how differences between science mapping approaches affect the results obtained and how the results can best be interpreted. To develop a better understanding of different mapping approaches, researchers have applied these approaches to the same dataset and have compared the outcomes (Gläser et al., 2017; Velden et al., 2017a; Velden et al., 2017b; Boyack & Klavans, 2010; Klavans & Boyack, 2017; Waltman et al., 2020). These comparative studies encompass a wide range of approaches, including diverse citation-based methods and text-based methods (Van Eck & Waltman, 2017; Velden et al., 2017b; Boyack et al., 2011; Klavans & Boyack, 2017; Waltman et al., 2020).

Topic modeling is a frequently used text-based mapping approach, but there are few studies that systematically compare topic modeling to other mapping approaches. Boyack et al. (2011) compared the accuracy of nine text-based similarity approaches and included topic modeling in their comparison. Their research focused on the accuracy of clustering outputs obtained using topic modeling. They did not perform a more in-depth exploration of the characteristics of their topic modeling results. Developing a better understanding of the value of topic modeling in the science mapping context therefore remains an open problem. This point was also made by Velden et al. (2017a), who suggested that future comparative studies should consider topic modeling approaches and potentially hybrid techniques that combine document clustering and topic modeling.

To fill this gap, we compare topic modeling (TM) with citation-based clustering (CC) as introduced by Waltman and Van Eck (2012). TM maps a research field based on co-occurrences of words, while CC constructs science maps based on citation flows. Obvious differences exist between words and citations, and therefore TM and CC capture different aspects of the communication in a research field. The comparison of TM and CC contributes to developing a more robust understanding of different science mapping approaches.

TM encompasses various methods for structuring large amounts of text data (Daenekindt & Huisman, 2020), with Latent Dirichlet Allocation (LDA; Blei, 2012, Blei et al., 2003) being the most widely used method in the scientometric field. However, scientometricians still have only a limited understanding of the use of LDA for science mapping. In this paper, we focus on comparing LDA and CC approaches, presenting a systematic comparison of the results obtained using the two approaches.

We adopt the same perspective as Gläser et al. (2017)—there is not one “best way” of structuring a research field, and thus, there is no ground truth to compare results. Based on this perspective, we present a systematic comparative analysis of TM and CC to provide insight into their respective strengths, weaknesses, and the characteristics of the results they provide. We take cardiovascular research (CVR) as a case study. Specifically, we create two maps of CVR, a TM map and a CC map. We then perform cluster-to-topic and topic-to-cluster mapping based on these maps. Our work identifies four types of relations between topics and clusters: one-to-one, one-to-many, many-to-many, and unique entities identified exclusively by either TM or CC. These relations highlight both similarities and differences between TM and CC in how they structure the CVR field. Based on this, we discuss advantages and disadvantages of each method.

Our work addresses the following research questions:

- To what extent do the results provide by TM and CC exhibit similarities or differences, and what is the nature of these similarities or differences?
- How can scientometricians choose between TM and CC?
- What considerations should scientometricians bear in mind when interpreting results obtained using TM and CC?

The paper is organized as follows. Section 2 provides a brief overview of related research. In Section 3, we introduce our dataset and describe the methodology employed. Section 4 presents a detailed analysis of the results obtained using TM and CC, and a comparison that explores the relations between topics and clusters. Section 5 summarizes our conclusions and proposes potential avenues for future research.

## 2 Related works

### 2.1 Topic modeling in scientometric studies

Topic modeling (TM) involves a group of statistical methods to extract latent semantic information from large textual datasets. Topics are typically defined as latent semantic structures occurring within a given corpus, in the form of words and probabilities (Jin et al., 2022; Daenekindt & Huisman, 2020). Documents are seen as probability distributions over topics (Curiskis et al., 2020; Chen et al., 2021; Jung et al., 2020).

In six journals that publish many scientometric studies (i.e., *Quantitative Science Studies, Journal of the Association for Information Science and Technology*, *Journal of Informetrics*, *Scientometrics*, *Frontiers in Research Metrics and Analytics*, *Journal of Data and Information Science*), there are 278 articles in the period 2010-2024 that cite Blei et al. (2003), the paper in which LDA was introduced. As the most widely used topic model in scientometrics research, LDA has been applied extensively to detect topics and specialties in a research area (Qi et al., 2018; Lamba & Madhusudhan., 2019; Song & Suhment., 2019) and to study the intellectual structure of science.

Scientomaticians also applied various LDA extensions, such as Hierarchical Latent Dirichlet Allocation (Griffiths et al., 2004) (cited in 3 scientometric publications), Dynamic Topic Model (Blei & Lafferty, 2006) (cited in 30 scientometric publications), Topic Over Time model (Wang & McCallum, 2006) (cited in 30 scientometric publications).

LDA and its extensions use bag-of-word representations of documents as input, which ignores both word order and semantic relationships (Švaňa, 2023, Grootendorst, 2022, Angelov, 2020). Some researchers argued that bag-of-word representations may fail to represent documents accurately (Voskergian, Jayousi & Yousef, 2024; Grootendorst, 2022). In response to this, word and document embedding techniques have gained popularity and have been used as a basis for extracting topics, for instance using approaches such as BERTopic (Grootendorst, 2022) and Top2Vec (Angelov, 2020).

BERTopic and Top2Vec are state-of-the-art embedding-based topic models. They aim to capture the semantics of words or documents by relying on an embedding representation. Specially, BERTopic uses the sentence-BERT framework to convert documents to embedding vectors (Grootendorst, 2022), while Top2Vec leverages Word2Vec, Doc2Vec or Universal Sentence Encoder to jointly generate document and word embedding vectors (Angelov, 2020). Both approaches then perform dimensionality reduction by applying the Uniform Manifold Approximation and Projection (McInnes, Healy & Melville, 2020) and the Hierarchical Density-Based Spatial Clustering of Applications with Noise (McInnes, Healy & Astels, 2017) algorithms to optimize the clustering process. BERTopic and Top2Vec both automatically determine the number of topics and allow performing topic reduction by grouping similar topics hierarchically (Grootendorst, 2022; Angelov, 2020).

Although embedding methods hold promise for enhancing topic representation, the limitations of these approaches, especially in terms of the interpretability of topics (Ballester & Penner, 2022; Lamirel, Lareau & Malaterre, 2024; Mihajlov et al., 2024), need to be acknowledged as well. It has been reported in the literature that embedding-based TM approaches yield results that are more difficult to interpret than the results of LDA, the TM approach that we focus on in the present paper (Ballester & Penner, 2022; Lamirel, Lareau & Malaterre, 2024).

## 2.2 Citation-based clustering approaches in scientometric studies

In scientometric research, citation linkages appear as tracers of intellectual connections (Zitt et al., 2010). Numerous citation-based mapping methods have been developed to analyze the intellectual structure of research fields. The prevalent clustering methods grounded in citations are co-citation, bibliographic coupling and direct citation.

In a co-citation network, a link is established between two publications when they are jointly cited by a third publication. Therefore, co-citation is most suitable for clustering older papers, and excludes recent papers that have not yet been cited (Petrovich, 2020, Boyack & Richard, 2010). Conversely, bibliographic coupling links publications sharing at least one common reference. Both co-citation and bibliographic coupling are indicators of the relatedness of publications. Direct citation provides another indicator of the association of publications. Waltman and Van Eck (2012) proposed a new approach for clustering publications based on direct citation. They demonstrated the application of this approach by clustering 10 million papers based on 97 million direct citations between these papers. Klavans and Boyack (2017) compared the accuracy of topic-level taxonomies based on clustering of documents using direct citation, bibliographic coupling, and co-citation. They found that using cited references from papers with a larger number of references, direct citation is better at clustering documents than both bibliography coupling and co-citation.

Direct citation offers a more direct mechanism for indicating relatedness of publications compared to co-citation and bibliographic coupling (Waltman & Van Eck, 2012). The clustering approach based on direct citation introduced by Waltman and Van Eck (2012) has been frequently used in recent science mapping literature. We therefore focus on this approach in the present paper.

## 2.3 Comparison of science mapping approaches

Two fundamentally different perspectives can be found in the literature on comparing science mapping approaches. One perspective focuses on assessing the validity or accuracy of different science mapping approaches, assuming the existence of an absolute notion of accuracy (Klavans & Boyack, 2017). A key challenge is the lack of a generally accepted benchmark to evaluate the accuracy of the results obtained using different science mapping approaches. Kevin Boyack, Dick Klavans, and their colleagues proposed to use a more or less independent method as a benchmark to assess the accuracy of different science mapping approaches. They compared various citation-based approaches, such as co-citation analysis, bibliographic coupling, and direct citation (Boyack & Klavans, 2010; Klavans & Boyack, 2017), as well as different text-based approaches (Boyack et al., 2011, Boyack, Small, & Klavans, 2013). Waltman et al. (2020) built on the basic idea put forth by Boyack and Klavans. They introduced a more systematic methodology for comparing different measures of relatedness and evaluating the accuracy of clustering solutions obtained using various citation-based and text-based measures.

The alternative perspective taken in the literature starts from the idea that there is no single best approach for identifying topics or characterizing the intellectual structure of a research field. There may be different meaningful and valid perspectives on the structure of a field (Gläser et al., 2017). Several papers in a special issue of *Scientometrics* titled "Same data—different results" took this perspective to compare the results obtained by applying different science mapping approaches to the same dataset. Van Eck and Waltman (2017), Havemann et al. (2017), and Velden et al. (2017b) extracted a direct citation network from the dataset and applied different clustering algorithms to this network. Velden et al. (2017a) found that clustering solutions for a network of astrophysics papers linked by direct citations (Havemann et al., 2017) captured major topics identified by other approaches, along with new topics not detected by those approaches. These new topics effectively represent 'bridging' or 'emerging' topics, highlighting the potential for detecting emerging developments in the literature. Additionally, clustering solutions obtained by clustering a hybrid network based on a combination of bibliographic coupling and textual similarity (Glänzel & Thijs, 2017) did not show a strong similarity with other approaches. Instead, these solutions offered their own unique perspective on the data. This underscores the notion that different thematic structures emerging from different methodological approaches may all be meaningful and valid in their own right.

In our paper, we adopt the perspective that there is no single best approach to science mapping. Based on this perspective, we undertake a comparison of two widely-used approaches, namely TM and CC.

# 3 Data and methods

To address the research questions presented in Section 1, we performed a systematic comparison of the TM and CC science mapping approaches. The following subsections elaborate on the properties of the two approaches, the data collection, and the data analysis.

### 3.1 Properties of the TM and CC science mapping approaches

Table 1 summarizes the key properties of the TM and CC approaches to science mapping studied in this paper.

Table 1. Properties of the TM and CC science mapping approaches

| | TM approach | CC approach |
|---|---|---|
| Method | Latent Dirichlet Allocation (Blei et al., 2003) | Citation-based clustering (Waltman and van Eck, 2012) |
| Relations between documents | Shared words in titles and abstracts of documents | Direct citation links between documents |
| Parameters | Number of topics ($K$)<br>alpha ($\alpha$)<br>beta ($\beta$)<br>Minimum probability<br>Stopword list | Resolution<br>Minimum cluster size |
| Clustering | A document is represented by a probability distribution over topics | A document is assigned to a single cluster |
| Visualization method | Principal Component Analysis | VOS |
| Dataset | Local data | Global data |

The basic idea behind LDA is that each document is represented as a mixture of latent topics with different proportions (Blei et al. 2003; Blei, 2012). This idea aligns with the nature of real-world documents, where different topics appear in varying degrees (Blei, 2012). LDA assumes that documents about the same topic use similar words, and it generates topics based on the co-occurrence of these words. Each topic is represented as a distribution over words, with these distributions modeled independently of one another. The assumption of independence between topics simplifies the model and ensures computational efficiency. However, in practice, topics often overlap due to the shared words across topics, which introduces complexity beyond the independence assumption.

LDA requires users to define the number of topics ($K$) and other parameters such as alpha ($\alpha$), beta ($\beta$) and minimum probability. A critical issue in successfully applying LDA is choosing an appropriate $K$. If $K$ is too low, topics become overly broad; if it is too high, "over-clustering" results in many small high-similar topics (Greene et al., 2014). The Dirichlet prior parameters $\alpha$ and $\beta$ influence the document-topic distribution and topic-word distribution respectively. A higher $\alpha$ broadens the topic coverage of a document, while a lower $\alpha$ concentrates a document on fewer topics. Similarly, a higher $\beta$ produces a more uniform word distribution for a topic, whereas a lower $\beta$ concentrates a topic on fewer words. In practice, a widely used approach to determine suitable parameter values and to evaluate robustness across parameter values is to experiment with a variety of different parameter values.

Researchers tend to cite papers within specific scientific communities or disciplines. CC clusters publications based on their relatedness represented by direct citations. CC assigns each document to a single cluster, yielding clusters with clear-cut boundaries. Unlike LDA, which requires specifying the number of topics in advance, CC does not expect the number of clusters to be determined in advance. Instead, CC determines the number of clusters based on two parameters: the resolution and the minimum cluster size. Higher resolution leads to more clusters, while lower resolution results in fewer clusters. Increasing the minimum cluster size reduces the number of clusters by merging smaller clusters.

To compare science mapping approaches based on LDA and CC, we employed widely used methods for visualizing topics and clusters. We used Principal Component Analysis (PCA) to visualize topics, while the VOS method was used to visualize clusters. We used these visualization methods because they are embedded in two well-established tools, LDAvis and VOSviewer. PCA is the default method for scaling high-dimensional topic distributions into a two-dimensional space (Sievert & Shirley, 2014). The popular VOSviewer tool employs the VOS method for constructing bibliometric maps, aiming to locate items (e.g., clusters of documents) in a two-dimensional space in such a way that their relatedness is represented as accurately as possible (Van Eck et al., 2010).

### 3.2 Data collection

Data was collected from the WoS database, specifically from three citation indexes: Science Citation Index Expanded, Social Sciences Citation Index, and the Arts & Humanities Citation Index. We used the in-house version of the WoS database, which was updated until the 13th week of 2020, at the Centre for Science and Technology Studies (CWTS) at Leiden University.

We constructed a dataset focusing on publications related to CVR using a hybrid information retrieval technique that integrates lexical and citation-based methods. This retrieval strategy is derived from Gal et al. (2015). The dataset encompasses articles, reviews, and letters that were published between 2010 and 2020. The dataset consists of 433,642 documents.

### 3.3 Data analysis

#### 3.3.1 Constructing and interpreting the TM map

TM encompasses a group of methods that aim to effectively structure extensive text data by identifying patterns of co-occurrences of terms within similar texts (Daenekindt & Huisman, 2020). In recent years, a variety of TM methodologies have emerged, with LDA being the most renowned and widely adopted method (Jockers & Thalken, 2020). We trained an LDA model using the Gensim library to analyze the titles and abstracts of the retrieved publications. The approach we utilized can be subdivided into two steps, which are detailed below.

*Step 1: Constructing the TM map*

It is essential to perform text preprocessing prior to training the LDA model. We employed natural language processing (NLP) tools to clean up the titles and abstracts of the retrieved documents. We applied the maximal matching algorithm, which relies on a dictionary created from the MeSH tree and expert knowledge, to extract field-specific n-grams from the titles and abstracts. After extracting the n-grams, nouns were extracted from the remaining text. We first applied standard preprocessing techniques. These techniques involved addressing typographical errors, tokenizing the text data while excluding specific phrases, lowercasing words, removing punctuation marks and special characters like numbers and URLs, removing stopwords, lemmatizing the text, and conducting part-of-speech (POS) tagging. After this preprocessing process, we extracted the nouns, where a noun was defined as a word with the POS tag “NNP”, “NN”, or “NNS”.

To optimize the performance of the LDA model, we excluded frequently occurring n-grams and nouns. We removed all n-grams and nouns that occurred in at least 95% of all documents or that were among the 100 most frequently occurring n-grams and nouns. This resulted in a refined dictionary consisting of 20,930 n-grams and nouns.

Setting the hyperparameters of the LDA model is a delicate and crucial process. We followed a typical approach to determining these hyperparameters, which involves performing a series of sensitivity tests. The hyperparameters are the number of topics k and the Dirichlet hyperparameters $\alpha$ and $\beta$. We conducted a series of sensitivity tests to systematically adjust these parameters and observe the effects on the results obtained from the LDA model. We conducted extensive testing using a variety of parameter combinations to evaluate different results. Specifically, we examined the topic extraction process across a range of topic counts, with 10, 15, 20, 25, 30, 40, 50, 100, 150 and 200 topics, and for two values of $\beta$, 0.01 and 0.1. For each parameter combination, an expert scrutinized the resulting topics. The expert, Qiao Zhao, is affiliated with the Leiden University Medical Center (LUMC) and possesses over 5 years of experience in the field. He leveraged his extensive domain knowledge in the field of CVR to assess the results obtained for different parameter combinations.

Ultimately, the parameters were set as follows: $\alpha = 1.0/k$, $\beta = 0.1$, and $k = 40$. Additionally, we performed Gibbs sampling for 5000 iterations to analyze the corpus. Subsequently, we trained the final model to obtain a term-topic distribution and document-topic distribution. We utilized PyLDAvis to visualize the topics. The resulting visualization is presented in Figure 2. To explore the TM map interactively, please download and open the HTML file available at https://bit.ly/44HZufm.

*Step 2: Interpreting the TM map*

In order to better understand the TM map, we labeled the topics. The labels assigned to each topic were not simply derived algorithmically from terms generated by the LDA model. Such an algorithmic approach may not accurately capture the essence of a topic. Moreover, in our research, the terms provided by LDA are more useful for describing specific issues than for capturing broad topics. Broad topics necessitate generic terms, such as "*clinical study*", "*cell study*", or "*pathology study*", which are frequently absent from journal titles and abstracts. To address this challenge, we employed a manual labeling strategy that involved a thorough examination of the top 40 terms, top 20 titles, and MeSH tree for each topic. The process of labeling topics encompassed three steps:

1. Obtaining MeSH tree labels for the terms within each topic. We identified terms in the MeSH tree corresponding to the terms in a topic. We started at the level where terms are, and then moved up levels until the highest level where the corresponding root nodes are located. In the end, we established the hierarchical structure of MeSH tree labels for every topic.
2. Selection of top 20 most relevant publications for each topic. Publications were selected based on their probability of belonging to a particular topic. We selected the top 20 publications with the highest probability.
3. Identification of the most appropriate labels. In the final step, topic labels were determined manually by analyzing the top 40 terms, top 20 publication titles, and the hierarchical MeSH tree labels. An illustration of a topic can be seen in Figure 2. The topic is characterized by frequent terms such as "*heart failure*", "*ejection*", "*fraction*", and "*hfpef*". The majority of the publication titles focus on the prognosis of heart failure and its analysis. Moreover, the MeSH tree labels are "Cardiovascular disease—heart disease—heart failure", and "Cardiovascular disease—Heart disease—Heart failure–Heart Failure, Diastolic". Based on the aforementioned findings and in collaboration with domain experts, we labeled the topic "Cardiovascular disease—heart disease—heart failure".

### 3.3.2 Constructing and interpreting the CC map

The CC method proposed by Waltman and Van Eck (2012) clusters publications based on direct citation relations, resulting in a hierarchical classification system with three levels. We utilized the most up-to-date hierarchical classification system created at CWTS using the method of Waltman and Van Eck in conjunction with the Leiden algorithm for network clustering (Traag, Waltman & van Eck, 2019). The classification system comprises 24 macro-level research areas, 812 meso-level research areas, and 4,140 micro-level research areas. Each publication is assigned to a single research area at the micro level of the classification system (Waltman & Van Eck, 2012). We used VOSviewer to build and visualize a CC map. Our strategy involved two steps.

*Step 1: Constructing the CC map*

For each micro-level research area, we determined the number of CVR publications and the total number of publications. We then calculated the percentage of CVR publications in each area by dividing the number of CVR publications by the total number of publications in the area. We selected the 142 clusters with at least 10% CVR publications. This selection was based on a critical examination of the titles of the top 10 most cited publications, the top 10 most frequent terms based on their absolute frequency, and the MeSH tree (as elaborated in step 2 of our methodology). We found that clusters with less than 10% CVR publications were not closely aligned with the CVR field, and their inclusion could introduce significant noise into the science map.

We used the VOSviewer software to visualize the 142 selected clusters. The resulting CC map can be accessed at https://bit.ly/3Aa2hzo. To determine suitable parameter settings for the CC map, we performed a series of analyses for different values of the resolution parameter. We explored values ranging from 0.7 to 1.5 with a step size of 0.1. We scrutinized the clusters obtained for each parameter value. Based on this, we decided to set the resolution parameter to 0.9. The cluster size parameter was set to 10. This yielded a map with three categories of clusters.

*Step 2: Interpreting the CC map*

In this step, we conducted an interpretation of the CC map by providing labels for every cluster. We did not, however, follow the labeling strategy suggested by Waltman and van Eck (2012), which aims to identify the top 5 terms that are most characteristic of a cluster. Instead, we manually assigned a label to each cluster through a comprehensive examination of the titles of the top 10 cited publications, the top 10 most frequent terms based on absolute frequency, and the MeSH tree. This manual labeling process allowed us to provide a detailed description of each cluster, and also of the three categories in which we grouped the clusters. Our approach for labeling clusters and identifying three categories involves the following three stages:

1. Obtaining MeSH tree labels. The procedure is similar to the procedure in step 1 of the TM labeling process. We identified terms in the MeSH tree corresponding to the terms in a cluster. We started at the level where terms are, and then moved up levels until the highest level where the corresponding root nodes are located. In the end, we established the hierarchical structure of MeSH tree labels for every cluster.
2. Selection of the top 10 most relevant publications. In the second stage, we counted the number of citations received by each paper in a cluster. We then identified the top 10 articles with the highest citation counts.
3. Determining a suitable label. In the final step, we determined a suitable label for a cluster by examining the top 10 most frequent terms, the titles of the top 10 most cited publications, and the hierarchical MeSH tree labels. Figure 3 provides an example of a cluster. For this cluster, the MeSH tree labels are "Cardiovascular disease—heart disease—heart failure" and "Cardiovascular disease—Heart disease—Heart failure–Heart Failure, Diastolic". The most frequent terms are "*heart failure*", "*prognosis*", "*acute heart failure*", "*chronic heart failure*", and "*heart failure with preserved ejection*". The majority of the paper titles in this cluster focus on recommendations for identifying and treating heart failure. Based on these findings combined with input from domain experts, we labeled the cluster "Cardiovascular disease—heart disease—heart failure".

### 3.3.3 Identifying relations between topics and clusters

To systematically analyze the similarities and differences between the TM map and the CC map, we created a cluster-to-topic mapping and a topic-to-cluster mapping. The cluster-to-topic and topic-to-cluster mappings were obtained in the following steps.

*Step 1: Getting the document-to-topic matrix.* In step 1 in Section 3.2.1, we obtained a document-to-topic matrix. We use $Q_{it}$ to denote the probability that document *i* pertains to topic *t*.

*Step 2: Getting the document-to-cluster matrix.* We obtained a document-to-cluster matrix in step 1 in Section 3.2.2. We use $R_{ic}$ to indicate whether document *i* belongs to cluster *c* ($R_{ic}$=1) or not ($R_{ic}$=0).

*Step 3: Getting the cluster-to-topic mapping and the topic-to-cluster mapping.* We created these mappings based on the aforementioned two matrices. For each topic *t* and cluster *c*, we determined the probability $P_{ct}$ that documents in cluster *c* are in topic *t*. Using *n* to denote the total number of documents, $P_{ct}$ is given by

$$P_{ct} = \frac{\sum_{i=1}^{n} Q_{it} R_{ic}}{\sum_{i=1}^{n} R_{ic}}$$

For each cluster *c* and topic *t*, we also calculated the probability $P_{tc}$ that documents in topic *t* are in cluster *c*. $P_{tc}$ is given by

$$P_{tc} = \frac{\sum_{i=1}^{n} Q_{it} R_{ic}}{\sum_{i=1}^{n} Q_{it}}$$

Based on $P_{ct}$ and $P_{tc}$ , we established binary relations between clusters and topics. We considered a topic *t* and a cluster *c* to be related if $P_{ct}$ or $P_{tc}$ is greater than a specific threshold. Otherwise, they were deemed unrelated. We then visualized the relations using the *igraph* library in Python. Through the visualization, we gained insights into similarities and differences between the results obtained from TM and CC.

Figure 1 illustrates four different types of relations between topics and clusters. Firstly, there are instances where a topic and a cluster are related to each other while they have no relations with other topics or clusters (referred to as a one-to-one relation; illustrated in panel A in Figure 1). Secondly, there are situations where a single topic is related to multiple clusters, or conversely, a single cluster is related to multiple topics (referred to as a one-to-many relation; illustrated in panel B in Figure 1). Additionally, there are more complex scenarios where a cluster is related to multiple topics, and some of these topics are related to multiple clusters (referred to as a many-to-many relation; illustrated in panel C in Figure 1). Lastly, there are cases where a topic has no relations with clusters, or the other way around, a cluster has no relations with topics (referred to as a unique topic or cluster; illustrated in panel D in Figure 1).

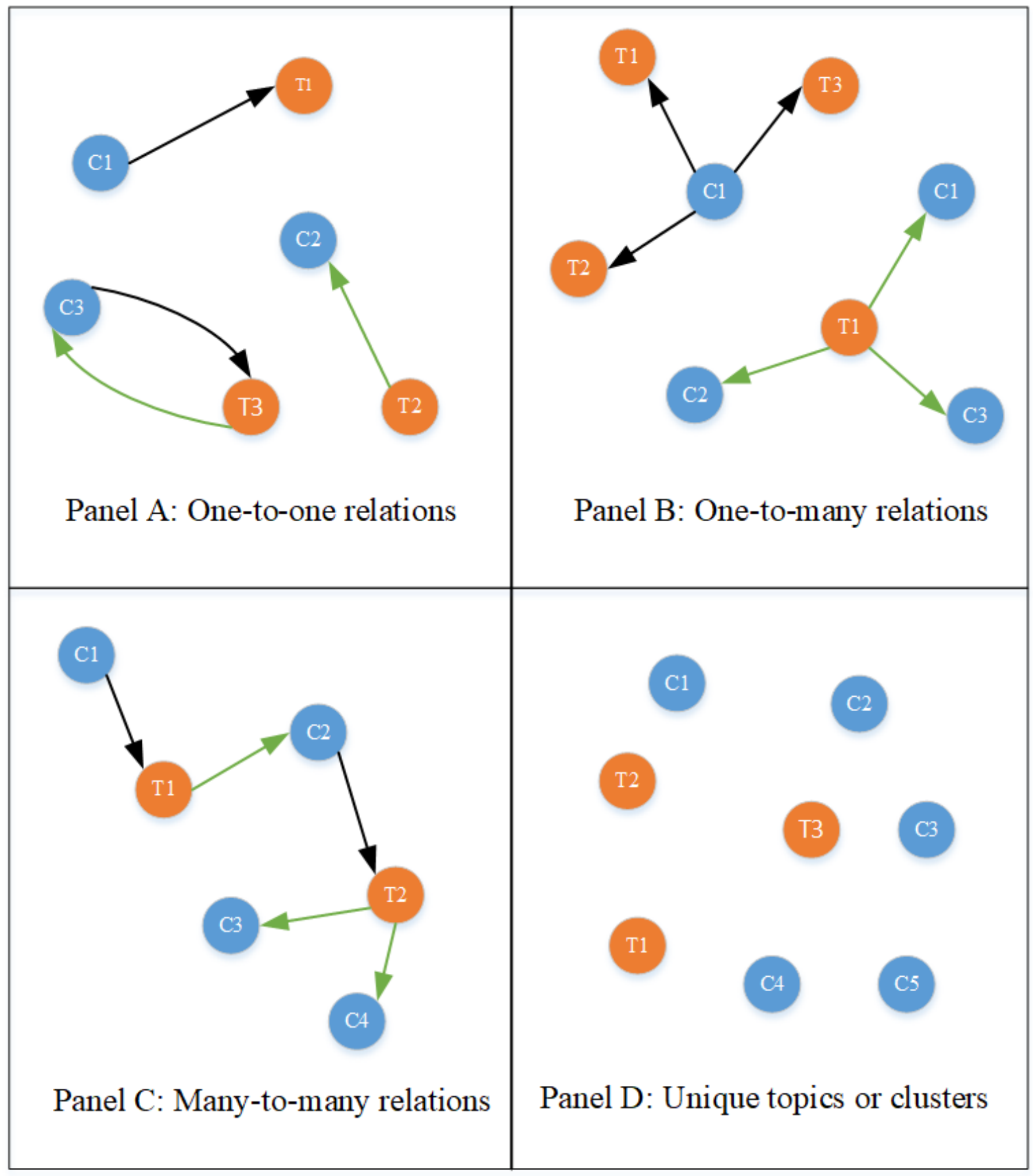

Figure 1. Four types of relations between topics and clusters

## 4 Results

### *4.1 Cardiovascular research through the lens of TM*

Figure 2 illustrates research landscape of CVR through the lens of TM. The plot consists of circles that indicate topics. The distance between two circles approximately signifies the degree of relatedness between topics. The size of the circles reflects the overall prevalence of each topic. The horizontal axis (PC1) in Figure 2 represents the distinction between clinical practice and physiological research. On the left side, topics pertaining to clinical trials and surgical procedures are represented, while the right side predominantly encompasses topics related to physiological research, such as the studies of cells and tissues. The vertical axis (PC2) reflects the diagnosis or therapy as well as the molecular composition of topics. The top portion of the axis corresponds to diagnostic techniques, gradually transitioning to surgical therapy at the bottom.

Figure 2 not only depicts topics in CVR but also visualizes the interconnections between topics based on semantic information. Consequently, we provide a comprehensive explanation of the map from both a content-oriented and structural standpoint. Based on the map, we categorized CVR into three main areas by examining the key terms for each topic and the topic labels (see step 2 in Section 3.2.1), supplemented by expert knowledge (Sievert et al., 2014): 1) *Physiological Studies* represented by the red category, 2) *Clinical Studies & Surgical Procedures* represented by the green category, and 3) *Risk Factors & Diagnosis Techniques* represented by the blue category. In more detail, for example, as shown in Figure A1 in the Appendix, the term 'risk factor' predominantly appears in topics located in the upper-left region of the map. With expert interpretation, these topics are classified into the category *Risk Factors.*

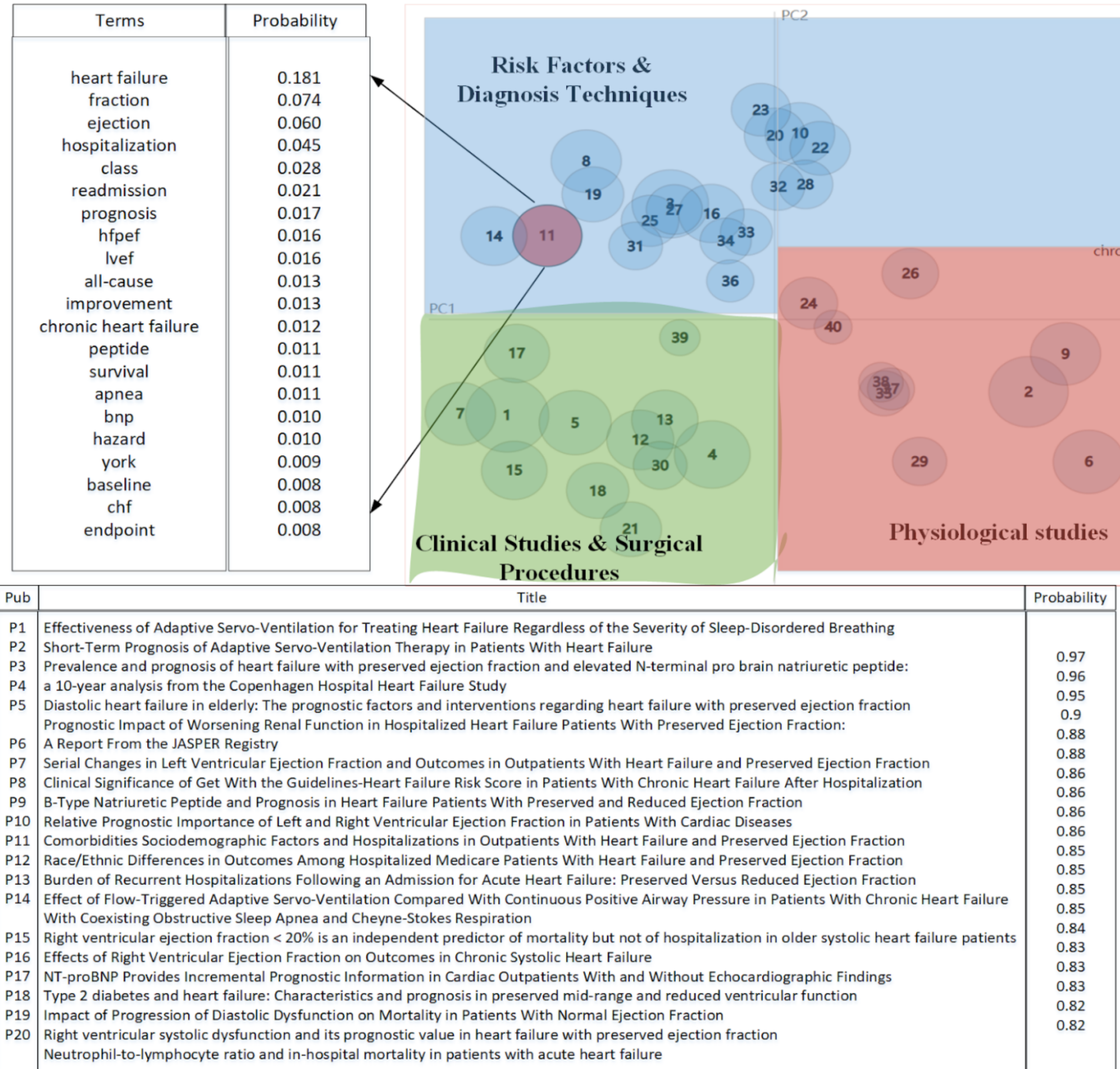

| Terms | Probability |
|---|---|
| heart failure | 0.181 |
| fraction | 0.074 |
| ejection | 0.060 |
| hospitalization | 0.045 |
| class | 0.028 |
| readmission | 0.021 |
| prognosis | 0.017 |
| hfpef | 0.016 |
| lvef | 0.016 |
| all-cause | 0.013 |
| improvement | 0.013 |
| chronic heart failure | 0.012 |
| peptide | 0.011 |
| survival | 0.011 |
| apnea | 0.011 |
| bnp | 0.010 |
| hazard | 0.010 |
| york | 0.009 |
| baseline | 0.008 |
| chf | 0.008 |
| endpoint | 0.008 |

| Pub | Title | Probability |
|---|---|---|
| P1 | Effectiveness of Adaptive Servo-Ventilation for Treating Heart Failure Regardless of the Severity of Sleep-Disordered Breathing | |
| P2 | Short-Term Prognosis of Adaptive Servo-Ventilation Therapy in Patients With Heart Failure | |
| P3 | Prevalence and prognosis of heart failure with preserved ejection fraction and elevated N-terminal pro brain natriuretic peptide: | 0.97 |
| P4 | a 10-year analysis from the Copenhagen Hospital Heart Failure Study | 0.96 |
| P5 | Diastolic heart failure in elderly: The prognostic factors and interventions regarding heart failure with preserved ejection fraction | 0.95 |
| | Prognostic Impact of Worsening Renal Function in Hospitalized Heart Failure Patients With Preserved Ejection Fraction: | 0.9 |
| P6 | A Report From the JASPER Registry | 0.88 |
| P7 | Serial Changes in Left Ventricular Ejection Fraction and Outcomes in Outpatients With Heart Failure and Preserved Ejection Fraction | 0.88 |
| P8 | Clinical Significance of Get With the Guidelines-Heart Failure Risk Score in Patients With Chronic Heart Failure After Hospitalization | 0.86 |
| P9 | B-Type Natriuretic Peptide and Prognosis in Heart Failure Patients With Preserved and Reduced Ejection Fraction | 0.86 |
| P10 | Relative Prognostic Importance of Left and Right Ventricular Ejection Fraction in Patients With Cardiac Diseases | 0.86 |
| P11 | Comorbidities Sociodemographic Factors and Hospitalizations in Outpatients With Heart Failure and Preserved Ejection Fraction | 0.86 |
| P12 | Race/Ethnic Differences in Outcomes Among Hospitalized Medicare Patients With Heart Failure and Preserved Ejection Fraction | 0.85 |
| P13 | Burden of Recurrent Hospitalizations Following an Admission for Acute Heart Failure: Preserved Versus Reduced Ejection Fraction | 0.85 |
| P14 | Effect of Flow-Triggered Adaptive Servo-Ventilation Compared With Continuous Positive Airway Pressure in Patients With Chronic Heart Failure | 0.85 |
| | With Coexisting Obstructive Sleep Apnea and Cheyne-Stokes Respiration | 0.85 |
| P15 | Right ventricular ejection fraction < 20% is an independent predictor of mortality but not of hospitalization in older systolic heart failure patients | 0.84 |
| P16 | Effects of Right Ventricular Ejection Fraction on Outcomes in Chronic Systolic Heart Failure | 0.83 |
| P17 | NT-proBNP Provides Incremental Prognostic Information in Cardiac Outpatients With and Without Echocardiographic Findings | 0.83 |
| P18 | Type 2 diabetes and heart failure: Characteristics and prognosis in preserved mid-range and reduced ventricular function | 0.83 |
| P19 | Impact of Progression of Diastolic Dysfunction on Mortality in Patients With Normal Ejection Fraction | 0.82 |
| P20 | Right ventricular systolic dysfunction and its prognostic value in heart failure with preserved ejection fraction | 0.82 |
| | Neutrophil-to-lymphocyte ratio and in-hospital mortality in patients with acute heart failure | |

Figure 2. TM map of CVR (https://bit.ly/44HZufm)

Within the category of *Physiological Studies*, we identified nine topics: *Regenerative Medicine* (T2), *Gene Transcription* (T6), *Oxidative Stress (T9)*, *Angiotensin (T29)*, *Genetic Research* (T24), *Antiplatelet Therapy (T35), Cell Studies-Cardiomyocytes (T26), Perfusion Re-injury (T37), and Cardiac Electrophysiology-Ion Channels (T38).* According to the size of topics, *Regenerative Medicine* holds the greatest prominence among the various topics within *Physiological Studies*.

In *Clinical Studies & Surgical Procedures*, there are ten topics related to the treatment of diseases through various approaches such as clinical medication, interventional methods, or surgical procedures. To elaborate further, T1 focuses on *Clinical*

*Guidelines for Medication*. T18 specifically addresses the use of *Anticoagulant Medication Treatment*. Within the realm of surgical procedures, T12 involves the *Surgical Treatment of Congenital Heart Disease,* while T39 discusses *Heart Transplantation*. Invasive treatments are covered in *Treatment of Arrhythmia* (T13), *Invasive Treatment of Myocardial Infarction* (T15), and *Interventional Therapy* (T21). Additionally, the topics of *Myocardial Ischemia* (T17) and *Life Assist Device* (T30) are included in this category. Taking into account the prevalence of these topics, it is evident that in the category of *Clinical Studies & Surgical Procedures* researchers are placing significant emphasis on *Clinical Guidelines for Medication*.

In terms of *Risk Factors & Diagnosis Techniques*, our focus revolves around the identification of causal factors and diagnostic indicators associated with symptoms, alongside an exploration of the diagnostic techniques. It is widely acknowledged that cardiovascular conditions stem from a combination of socio-economic, behavioral, and environmental risk factors, which we also discovered in our research. The risk factors include *Hypertension* (T3), *Behavioral Risk Factors* (T8) such as "diabetes", "obesity", "alcohol" and "tobacco use", *Socio-economic Risk Factors* (T31) encompassing "life quality", "anxiety", "depression", and "emotions", and *Cholesterol Level* (T10). There is also a topic on the *Ethnic Background Studies of CVR* (T14), which is related to studying risk factors related to race and gender. Risk factors are also studied for *Kidney Disease* (T34). Regarding the diagnostic indicators of symptoms, researchers study *Biomarkers* (T28) and *Heart Rate Variability* (T32). In addition, the following diagnostic techniques are studied: *Electrocardiogram* (T33), *Coronary Angiography* (T25) and *Magnetic Resonance Imaging* (T16). Based on the prevalence of topics, *Hypertension* and *Behavioral Risk Factors* have garnered heightened attention. In other words, research pertaining to risk factors has received considerable emphasis in this field of study.

In addition, the TM map provides valuable insights not only into the research content of CVR studies but also depicts the interconnectedness among the three categories. *Physiological Studies* exhibits close associations with *Risk Factors & Diagnosis Techniques*, whereas it demonstrates a more distant relationship with *Cardiovascular Diseases & Surgical Procedures*. This observation suggests a gap between clinical research and physiological studies.

In terms of *Physiological Studies*, *Regenerative Medicine* (T2) and *Oxidative Stress* (T9) exhibit interconnectedness. This is due to the fact that regenerative therapies, such as stem cell therapy, possess the potential to repair and regenerate damaged tissues, thereby mitigating oxidative stress and preventing further cellular damage. Conversely, oxidative stress can also impact the efficacy of regenerative therapies. Therefore, these two topics demonstrate a strong relationship, as their interactions are mutually influential. In *Risk Factors & Diagnosis Techniques*, there is a close association between *Heart Failure* (T11), *Ethnic Background Studies of CVR* (T14), *Behavioral Risk Factors* (T8) and *Hypertension* (T3). This connection arises from the fact that nearly all risk factors eventually lead to heart failure. Moreover, certain diagnostic indicators of risk factors necessitate the use of measurement instruments for accurate diagnosis. Consequently, a substantial correlation is observed between risk factors and diagnostic techniques. In *Clinical Studies & Surgical Procedures*, topics demonstrate significant interrelationships. This can be attributed to the critical role that clinical research plays in the evaluation of the safety and efficacy of surgical procedures.

Additionally, clinical studies serve as a crucial foundation for informing decisions about the best course of treatment for patients.

### *4.2 Cardiovascular research through the lens of CC*

The CC map of CVR is presented in Figure 3. Each circle in Figure 3 indicates a micro-level research area, while the proximity between two circles approximately represents the degree of relatedness based on direct citation links. The color of the circles represents groups of highly related research areas, and the size of the circles reflects the number of publications in a research area.

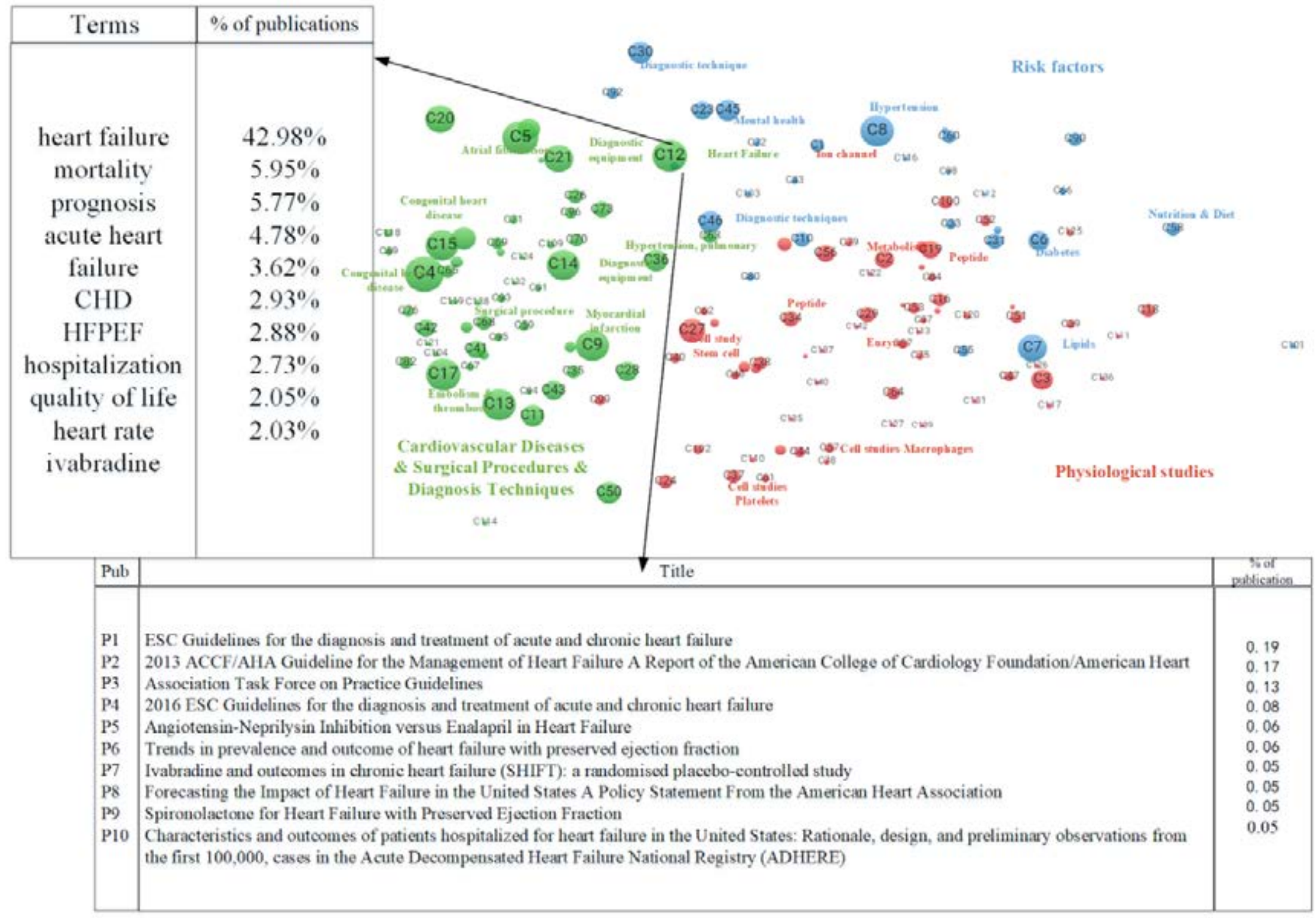

| Terms | % of publications |
|---|---|
| heart failure | 42.98% |
| mortality | 5.95% |
| prognosis | 5.77% |
| acute heart | 4.78% |
| failure | 3.62% |
| CHD | 2.93% |
| HFPEF | 2.88% |
| hospitalization | 2.73% |
| quality of life | 2.05% |
| heart rate | 2.03% |
| ivabradine | |

| Pub | Title | % of publication |
|---|---|---|
| P1 | ESC Guidelines for the diagnosis and treatment of acute and chronic heart failure | 0. 19 |
| P2 | 2013 ACCF/AHA Guideline for the Management of Heart Failure A Report of the American College of Cardiology Foundation/American Heart | 0. 17 |
| P3 | Association Task Force on Practice Guidelines | 0. 13 |
| P4 | 2016 ESC Guidelines for the diagnosis and treatment of acute and chronic heart failure | 0. 08 |
| P5 | Angiotensin-Neprilysin Inhibition versus Enalapril in Heart Failure | 0. 06 |
| P6 | Trends in prevalence and outcome of heart failure with preserved ejection fraction | 0. 06 |
| P7 | Ivabradine and outcomes in chronic heart failure (SHIFT): a randomised placebo-controlled study | 0. 05 |
| P8 | Forecasting the Impact of Heart Failure in the United States A Policy Statement From the American Heart Association | 0. 05 |
| P9 | Spironolactone for Heart Failure with Preserved Ejection Fraction | 0. 05 |
| P10 | Characteristics and outcomes of patients hospitalized for heart failure in the United States: Rationale, design, and preliminary observations from the first 100,000, cases in the Acute Decompensated Heart Failure National Registry (ADHERE) | 0.05 |

Figure 3. CC map of CVR (https://bit.ly/3Aa2hzo)

From the overall view, Figure 3 illustrates that CC reveals three primary categories of research: 1) *Physiological Studies* represented by the red category, 2) *Cardiovascular Diseases & Surgical Procedures & Diagnosis Techniques* denoted by the green category, and 3) *Risk Factors* identified by the blue category.

In *Physiological Studies*, we identified various levels of physiological investigations, encompassing *Cell Level Studies* (C27, C37, C57), *Gene Level Studies* (C16), *Hemodynamic Studies* (C24, C37, C44, C51, C81,), and *Ion Channel Level Studies* (C56). Additionally, our analysis highlights a concentration of research endeavors in the areas of *Enzyme Studies* (C24), *Peptide Studies* (C81), and *Protein Studies* (C44). In the end, our analysis reveals a great emphasis on publications pertaining to *Cell Level Studies* (C27, C37, C57), whereas *Gene Level Studies* (C16), *Hemodynamic Studies* (C24, C37, C44, C51, C81) and *Ion Channel Level Studies* (C56) exhibit relatively fewer scholarly contributions.

Within the realm of *Cardiovascular Diseases & Surgical Procedures & Diagnosis Techniques*, CC identifies four primary cardiovascular diseases referring to the MeSH tree: Cardiovascular Abnormalities, Cardiovascular Infractions, Heart Diseases, and Vascular Diseases. Delving into further details, Heart Diseases encompass various conditions, including *Atrial Fibrillation* (C5), *Heart Failure* (C12), *Cardiomyopathies* (C49), and *Heart Arrest* (C26). Vascular Diseases encompass conditions such as *Arterial Occlusive Diseases* (C95), *Aortic Aneurysm* (C41), *Embolism and Thrombosis* (C17), *Hypertension* (C8), *Pulmonary Hypertension* (C36), *Aneurysm Dissection* (C42), *Myocardial Ischemia* (C9), and *Varicose Veins* (C59). Additionally, CC identifies two types of surgical procedures, namely *Cardiac Surgical Procedures* and *Vascular Surgical Procedures*, exemplified by interventions such as *Coronary Artery Bypass Grafting* (C68), *Heart Valve Prosthesis Implantation* (C4), and *Percutaneous Coronary Intervention* (C82). Moreover, this map highlights *Diagnosis Techniques* (C20, C21, C73) as well. Based on the aforementioned analysis and clusters' size, it is evident that there are a great number of publications focused on cardiovascular diseases, particularly *Heart Failure* (C12), *Atrial Fibrillation* (C5), and *Myocardial Ischemia* (C9). This phenomenon reflects the focus of CVR on mainstream disease studies. Furthermore, the CC map reveals a multitude of clusters associated with disease, while surgical procedures and diagnostic techniques are represented by fewer clusters.

*Risk Factors* encompasses eight risk factors: *Hypertension* (C8), *Mental Health* (C45), *Climate Change* (C77), *Alcohol* (C66), *Diabetes* (C6), *HIV & AIDS* (C55), *Nutrition & Diet* (C18, C58), and *High Lipoprotein* (C7). These are important socio-economic, behavioral, and environmental risk factors. There is a strong concentration of publications centered around *Hypertension* (C8), *Diabetes* (C6) and *High Lipoprotein* (C7). Conversely, there is relatively less attention directed towards *Climate Change* (C77), *Mental Health* (C45), *Alcohol* (C66), and *Nutrition & Diet* (C18, C58). In addition to the aforementioned cardiovascular studies, CC also uncovers some small and specific clusters such as *Salty Food Intake* (C90), *Adiponectin* (C51), and *Lipid Breakdown* (C47).

Figure 3 illustrates that the CC map exhibits a similar relational structure to the TM map for CVR. Specifically, *Physiological Studies* exhibit a strong association with *Risk Factors* while displaying a comparatively weaker connection to *Cardiovascular Diseases & Surgical Procedures & Diagnosis Techniques*. Notably, *Risk Factors* constitute a distinct and discernible category within the CC map. Also, CC uncovers some specific clusters related to *Cardiovascular Diseases,* which are closely linked with *Surgical Procedures* and *Diagnosis Techniques.*

*Hemodynamic Studies* (C24, C44, C81), situated in the lower middle of the CC map within *Physiological Studies*, exhibit limited connections to other physiological studies. However, in the TM map, *Hemodynamic Studies* demonstrate a close association with *Diagnosis Techniques*. All clusters within *Risk Factors* are located at the top right of the CC map. *Cardiovascular Diseases & Surgical Procedures & Diagnosis Techniques* consist of interconnected clusters, with some clusters focusing on heart diseases (C4, C15, C25, C61, C65, C69, C107), some clusters on arterial disease (C14, C17, C41, C42, C59, C68, C86, C95), and others on venous diseases (C9, C11, C13, C28, C35, C43). This clear delineation of sub-structure highlights the categories within the domain of cardiovascular diseases.

### 4.3 Relations between topics and clusters

We constructed a cluster-to-topic and a topic-to-cluster mapping to further explore the relations between topics and clusters. The cluster-to-topic mapping provides the probability $P_{ct}$ of documents in cluster *c* belonging to topic *t*. Conversely, the topic-to-cluster mapping provides the probability $P_{tc}$ of documents in topic *t* belonging to cluster *c*. $P_{ct}$ and $P_{tc}$ offer different perspectives on the relatedness of topics and clusters. The consideration of both $P_{ct}$ and $P_{tc}$ enables a more comprehensive assessment of the similarity between topics and clusters.

$P_{ct}$ and $P_{tc}$ both range from 0 to 1, where values closer to 1 indicate stronger similarity, while a value of 0 implies no similarity at all between a topic and a cluster. It would be extremely challenging to analyze in full detail the similarities between all 40 topics and all 142 clusters. We therefore used a similarity threshold to simplify the investigation of relations between topics and clusters. We consider a topic *t* and a cluster *c* to be related if $P_{ct}$ or $P_{tc}$ is greater than a given threshold.

We manually reviewed the cluster-to-topic and topic-to-cluster mappings obtained using different thresholds. We utilized the *igraph* library in Python to visualize the mappings, as shown in Figures 4 and 5. The figures provide insights into three distinct categories of relations: one-to-one, one-to-many and many-to-many. For the sake of clarity, unique clusters or topics are not included in the visualizations. One-to-one relations signify a single cluster corresponding to a single topic and vice versa, as demonstrated in panel A of Figure 1. One-to-many relations refer to a single cluster that is associated with multiple topics, with each topic corresponding to only one cluster, or vice versa, as depicted in panel B of Figure 1. Many-to-many relations involve several clusters associated with various topics, as shown in panel C of Figure 1. In Figures 4 and 5, blue circles represent clusters, while orange circles indicate topics. The similarity between clusters and topics is represented by numerical values.

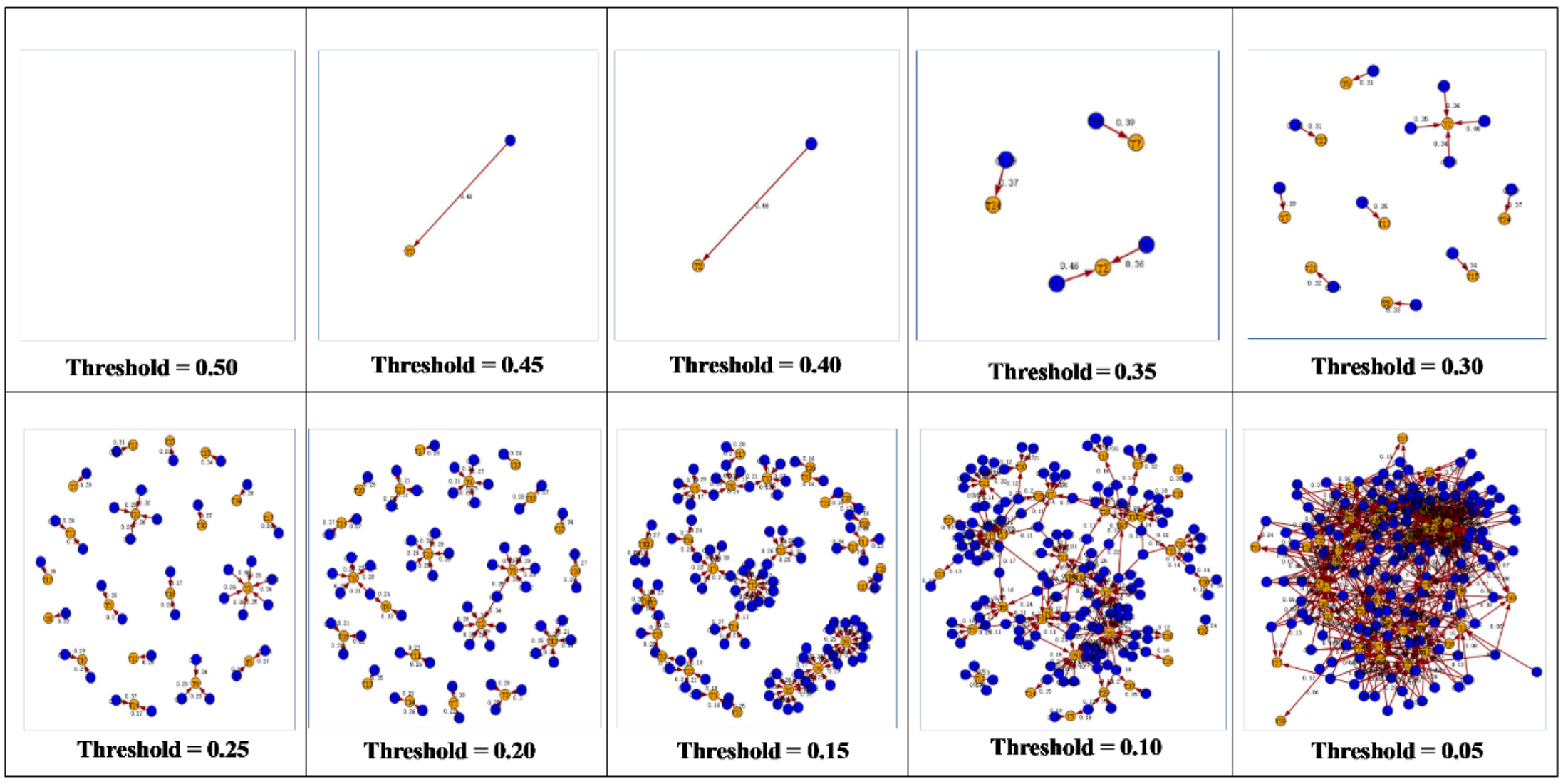

Figure 4. Cluster-to-topic relations for different thresholds

As evidenced in Figure 4, when the threshold for the cluster-to-topic mapping is set at 0.50, there are no relations between clusters and topics. With thresholds of 0.45 or 0.40, a single one-to-one relation is obtained between a cluster and a topic. Further reduction of the threshold to 0.35 reveals a combination of one-to-one and one-to-many relations between clusters and topics. Moreover, lowering the threshold to 0.30 or 0.25 uncovers additional one-to-one and one-to-many relations. When the threshold is set at 0.20, a diverse pattern emerges, including one-to-one, one-to-many and many-to-many relations. Notably, by lowering the threshold to 0.15, two larger many-to-many groups are formed. Reducing the threshold even more results in a higher density of connections between topics and clusters, predominantly characterized by many-to-many relations

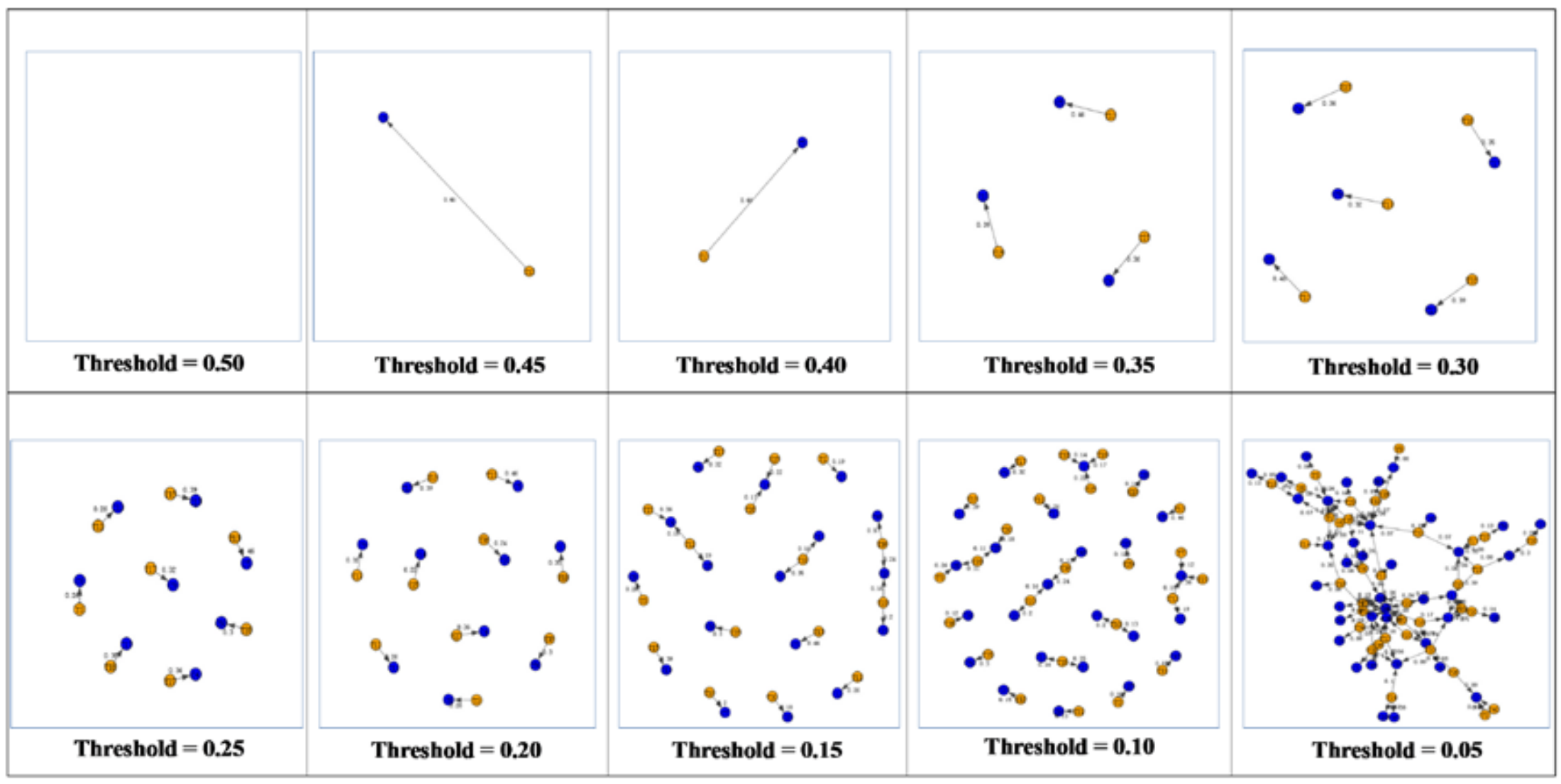

Figure 5. Topic-to-cluster relations for different thresholds

We now turn to the topic-to-cluster mapping. As evidenced in Figure 5, there are no relations between topics and clusters when the threshold is set to 0.50. With a threshold set at 0.45 or 0.40, a single topic is linked to a single cluster. As the threshold is further lowered to 0.35, 0.30, 0.25 or 0.20, there are three, five, eight, or ten pairs of a topic and a cluster, respectively, all exhibiting a one-to-one relation. When the threshold is set to 0.15, a mix of one-to-one, one-to-many, and many-to-many relations is obtained. Setting the threshold to 0.10 leads to a further increase in the relatedness of topics and clusters. Finally, when the threshold is reduced to 0.05, one large many-to-many group emerges.

Figures 4 and 5 reveal a notable absence of strongly related topics and clusters. Only in a few exceptional cases do more than one-third of the documents in a topic pertain to the same cluster, or vice versa. Consequently, relations between topics and clusters are generally relatively weak. In most cases, the overlap of documents between topics and clusters is less than 20%.

To gain deeper insights into the nature of the relations between topics and clusters, our investigation centers on relations that surpass specific thresholds: specifically, we consider all relations for which $P_{ct} \geq 0.2$ or $P_{tc} \geq 0.1$. At these thresholds, the data reveals different types of relationships (one-to-one, one-to-many, many-to-many) in a reasonably balanced way-(as illustrated in Figures 4 and 5). Additionally, we select a higher threshold for $P_{ct}$ than for $P_{tc}$ because on average the number of documents in a topic is larger than the number of documents in a cluster. Consequently, values of $P_{ct}$ can be expected to be greater than values of $P_{tc}$.

Figure 6 shows the relations between topics and clusters obtained using the above-mentioned thresholds. To improve clarity, unique clusters or topics are not included in the figure. The figure presents three types of relations: one-to-one, one-to-many and many-to-many. One-to-one and one-to-many relations are indicative of TM and CC identifying similar intellectual structures. Conversely, many-to-many relations and unique topics or clusters reveal differences in the intellectual structures identified by TM and CC.

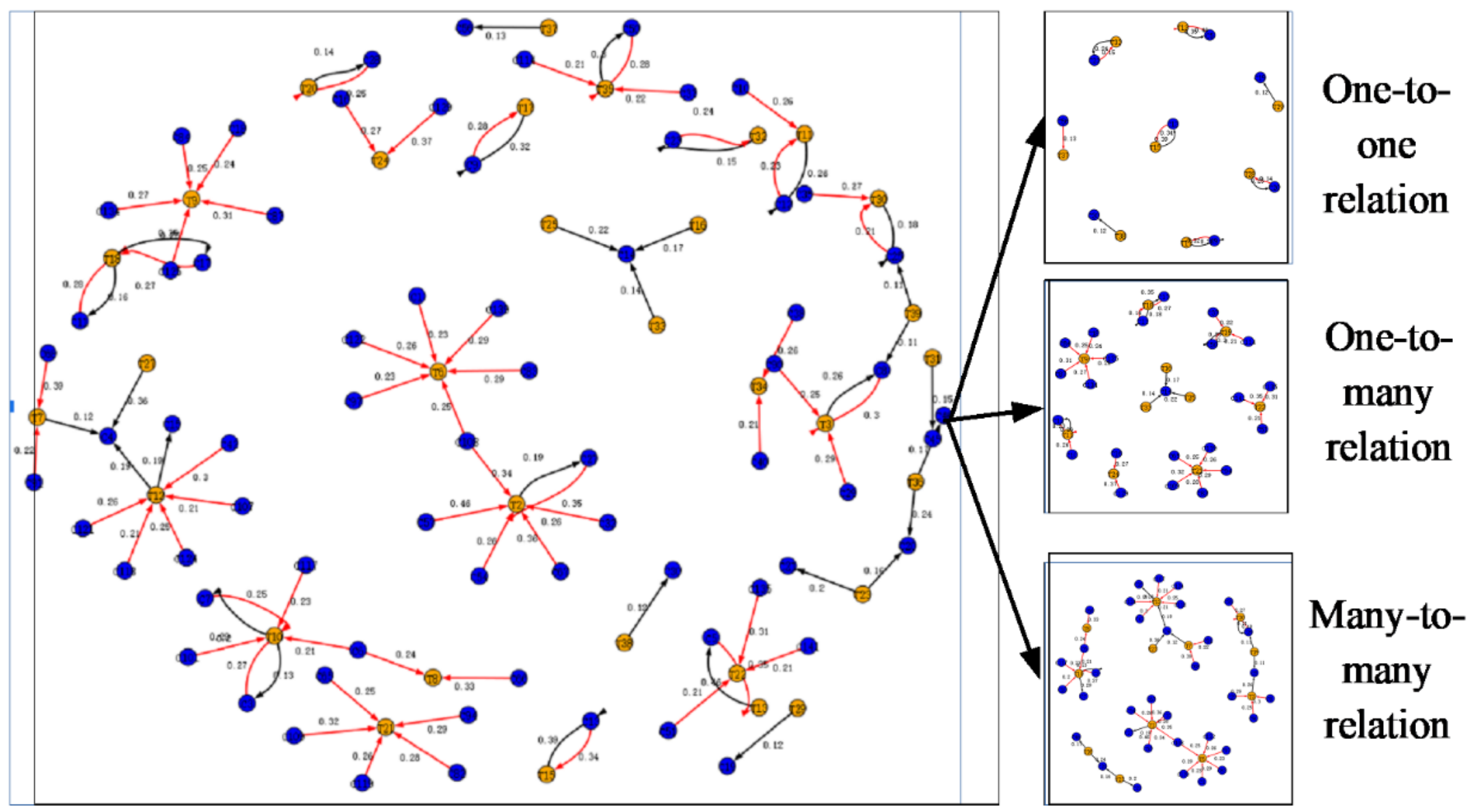

Figure 6. Relations between topics and clusters

From the nature of topics and clusters, both methods identified almost the same research areas within CVR. For example, both methods identified research areas such as *Cell Level Studies* (C27, C32, C54, C57, C67, C108, T2), *Gene Level Studies* (C16, C129, T24), *Biochemistry—Ion Channel Studies* (C56, T37), *Heart Failure* (C10, C12, T11), *Atrial Fibrillation–clinical studies* (C5, T13), Surgical Procedures of *Congenital Heart Disease* (C4, C15, C42, C107, C118, C121, C124, T12), and *Mental Health* (C45, T31). There are two kinds of relations that reveal similarities between topics and clusters, that is one-to-one and one-to-many relations.

In terms of one-to-one relation, the similarities are easily understood. The focus of our explanation will be on how similarities are revealed in one-to-many relations, as demonstrated in panel B of Figure 1. We observed that one cluster corresponds to several topics. For instance, C14 corresponds to T16, T25 and T33. To elaborate, C14 is associated with *Medical Imaging Techniques*, and includes terms such as "myocardial perfusion imaging", "computed tomography", "coronary computed tomography angiography", and "coronary angiography". Approximately 14% of the publications in *Electrocardiogram* (T33), 22% of the publications in *Coronary Angiography* (T25) and 17% of the publications in *Magnetic Resonance Imaging* (T16) are in *Medical Imaging Techniques* (C14). This implies that topics T16, T25 and T33 and cluster C14 identify a similar research area in CVR at different levels of granularity. TM provides a more refined classification compared to CC. We explored the underlying reasons contributing to the creation of one-to-many relations. CC categorizes publications that use similar materials, equipment, practical techniques, or tools used in the cited work, according to the citations of the methodological type proposed by Bornmann and Daniel (2008). As a result, CC tends to yield more generic results in the context of *Medical Imaging Techniques*. On the other hand, TM structures publications based on the co-occurrences of terms in similar texts (Daenekindt & Huisman, 2020). Consequently, TM distinguishes differences among diagnostic approaches employed for various diseases. In the case of *Medical Imaging Techniques,* several topics (e.g., T16, T25, T33) are associated with different aspects of the field. To sum up, regarding *Diagnosis Techniques,* clusters generated by CC provide a generic perspective on diagnostic techniques, while topics derived from TM depict specialized sub-techniques for diverse applications.

In the case of one-to-many relations, we observe oppositely corresponding relations, where one topic corresponds to multiple clusters, as illustrated in panel B of Figure 1. For instance, T21 is associated with clusters C65, C82, C94, C109 and C119. In more detail, T21 identifies terms such as “transcatheter”, “closure”, “catheter”, “PVI”, “catheterization”, and “reintervention”. It indicates an emphasis on *Interventional Treatment.* Approximately 25% of the publications in *Transcatheter Closure* (C65), 32% of the publications in *Transcatheter Closure-Pediatric research* (C109), 26% of the publications in *Vascular Surgery* (C119), 28% of the publications in *PVI* (C82) and 29% of the publications in *Hemodialysis Access* (C94) belong to *Interventional Treatment (T21)*. Although *Transcatheter Closure* (C65, C109), *Hemodialysis Access* (C94) and *PVI* (C82) all fall under the category of *Interventional Treatment*, their treatment objectives and principles differ. In a nutshell, regarding *Interventional Treatment*, topics generated by TM depict a generic perspective, while clusters obtained from CC offer a specialized classification for different treatment objects.

In terms of dissimilarities, there are two types of relations that manifest the dissimilarities. The first type is represented by a unique solution identified by either TM or CC, as depicted in panel D of Figure 1. As an example, TM discerns the unique topics *Practical Guidelines for Clinical Medication of CVR* (T1), *Prevention Strategies of Cardiovascular Diseases* (T4) and *Clinical Trial Studies* (T5). We explored the reasons for the existence of unique topics. We were aware that publications on these topics are distributed over several clusters, with a primary focus on medication adherence and risk factors. Additionally, some clusters, like *Nutrition & Diet* (C58), *Arterial Occlusive Diseases* (C95), *Food Chemical Elements* (C101), *Protein Studies of Biological Chemistry* (C113), and *Phlebology Studies* (C132) do not have corresponding topics. We examined the characteristics of these clusters and found that they contain a limited number of CVR publications, and these publications focus on several research objectives in one cluster. In short, TM groups publications into specific topics focused on *Practical Guidelines for Clinical Medication of CVR* (T1) and *CV Clinical Trial Studies* (T5), while publications within these topics are distributed among various clusters that center on different aspects of risk factors. Furthermore, CC generates clusters of small size that are characterized by their focus on several research objectives.

Many-to-many relations highlight the dissimilarities between TM and CC. Many-to-many relations refer to situations where a singular cluster is associated with multiple topics, and each topic is linked to multiple clusters, as exemplified in panel C of Figure 1. For instance, *Rheumatic Diseases* (C35) and *Life Assistance Devices and Heart Transplantation* (C25) display varying proportions of publications related to *Life*

*Assistance Devices* (T30). Furthermore, C25 has publications that are related to both *Life Assistance Devices* (T30) and *Heart Transplantation and Medication Adherence* (T39). This makes it a connecting point for both T30 and T39. Similarly, T39 serves as a bridge linking *Hypertension* (C8) and *Life Assistance Devices and Heart Transplantation* (C25). C8 also serves as a connection point linking *Heart Transplantation and Medication Adherence* (T39) and *Hypertension* (T3). These connection points link C25, C35, T30, T39, C8, and T3 together, forming a many-to-many relation. To delve deeper, 21% of publications in cluster *Life Assistance Devices and Heart Transplantation* (C25) and 27% of publications in cluster *Rheumatic Diseases* (C35) are in *Life Assistance Devices* (T30). And *Heart Transplantation and Medication Adherence* (T39) contains 11% of the publications in *Life Assistance Devices and Heart Transplantation* (C25) and *Hypertension* (C8). Meanwhile, T3 encompasses terms such as "hypertension", "preeclampsia", "blood pressure", "food", "salt" and so on. Furthermore, "preeclampsia" is a complication of pregnancy-induced hypertension, which is one subcategory of hypertension. "Food" and "salt" are the leading causes of hypertension. It shows that T3 focuses on *Hypertension.* Approximately 30% of the publications in *Hypertension* (C8) are in T3.

In summary, TM groups publications into topics such as *Hypertension* (T3), *Life Assistance Devices* (T30) and *Heart Transplantation* (T39), which depict interdisciplinary connections. In contrast, CC structures publications into clusters that center on different aspects of these topics, thereby creating a distinct division between *Risk Factors* and *Physiological Research*. In addition, we discovered topics that highlight the surgical or clinical research areas of CVR, with corresponding publications distributed across clusters that focus on specific diseases. Moreover, many-to-many relations demonstrate differences in intellectual structure. The TM map exhibits a close association between *Risk Factors* and *Diagnosis Techniques*, while the CC map reveals a strong connection between *Risk Factors* and *Diseases* or *Surgical Procedures.*

## 5 Discussion and conclusion

Various mapping approaches capture the intellectual structure of science at different levels and from different perspectives. Understanding these differences is crucial in determining how science maps can best be used to support decision-making processes in science. In this paper, we have presented a systematic comparison of the use of topic modeling (TM) and citation-based clustering (CC) for science mapping, aiming to gain a better understanding of the properties of these two approaches.

Our work aligns with previous studies that demonstrated both similarities and significant differences between different science mapping approaches (Velden et al., 2017a). Focusing on cardiovascular research (CVR) as a case study, we discerned four types of relations between topics and clusters, showing both similarities and differences between TM and CC: one-to-one, one-to-many, many-to-many, and unique entities identified exclusively by either TM or CC. The one-to-one and one-to-many relations signify overlap between topics and clusters, with TM and CC both identifying similar research areas within CVR. The many-to-many relations and unique entities reveal significant differences in the intellectual structure identified by TM and CC. Through an in-depth analysis of these relations, we analyzed the properties of CC and TM and developed strategies to enhance the interpretability of topics and clusters.

Our work reveals that relations between topics and clusters tend to be weak, with limited overlap of topics and clusters. Only in a few exceptional cases do more than one-third of the documents in a topic belong to the same cluster, or vice versa. Our work showcases both similarities and differences between TM and CC:

- CC performed effectively in identifying various diseases, whereas TM did not distinguish between disease topics. Topics generated by TM amalgamated disease terms, diagnostic techniques, risk factors, and treatment procedures terms.
- Both TM and CC recognized diagnostic techniques, but CC provided a holistic picture of diagnostic techniques, while TM identified sub-techniques that are distributed across multiple topics.
- In terms of *Clinical Treatment & Surgical Procedures,* both approaches identified similar content. However, they organized the same content in different ways. TM tended to describe generalized aspects, whereas CC produced more specialized clusters.
- TM classified specific topics related to *Practical Guidelines for Clinical Medication* and *Clinical Trial Studies of CVR*. In contrast, CC did not generate any specific clusters for these topics, instead organizing relevant publications into risk factor clusters.
- CC generated small clusters with a limited number of CVR publications. These clusters are indirectly related to CVR.
- In terms of intellectual structure, TM showcased a close link between *Risk Factors* and *Diagnosis Techniques*, while *Risk Factors* were closely connected to *Diseases* and *Surgical Procedures* in the CC map.

TM extracts semantic information from textual data, and groups publications based on co-occurrences of words. A document can be assigned to multiple topics. In some cases, TM is negatively affected by the presence of highly similar topics, making it difficult for users to observe clear differences between the topics. In our results, topics obtained using TM depicted specialized sub-techniques of diagnostic techniques for diverse applications. Most likely, this is because the sub-techniques diverge in their terminology in diagnostic techniques application. It aligns with the argument of Zitt et al. (2010), suggesting that in applied fields that heavily refer to a common theoretical or methodological substrate, the cited repertoire is almost identical but sub-communities diverge in their terminology. In this case, TM is likely to be more reliable to discriminate between areas of research, while CC connects theoretical aspects with various applications to form a single research area.

Moreover, various topics obtained using TM are strongly focused on the societal demand for cardiovascular disease prevention. For instance, some topics are concerned with unhealthy lifestyles and psychological stress as important risk factors for cardiovascular disease. On the other hand, clusters resulting from CC generally do not reflect societal needs. They primarily concentrate on understanding the physiological mechanisms of cardiovascular disease.

In contrast to TM, CC utilizes direct citations to group publications, with each document being assigned to a single cluster. Advantages of CC are simplicity and clear-cut clusters. In our analysis, CC managed to identify specific diseases as opposed to large disease clusters. In addition, CC produced specialized clusters focused on different treatments within the surgical and physiological research domains. Compared to TM, CC seems to more accurately capture differences in the intellectual structure of different diseases, such as pathological mechanisms, clinical manifestations and treatment methods. Consequently, CC has a stronger capability to delineate scientific micro-communities, and clusters obtained using CC provide a clearer picture of the intellectual structure of CVR.

We take the position that there is no "best method" in scientometrics. Every method has its own distinct characteristics, and these characteristics may be useful for some purposes and less useful for other purposes. Consequently, we advise users to carefully select suitable methods by considering the properties of different approaches and aligning them with specific research needs.

Our study also contributes to enhancing the interpretability of topics and clusters. We experienced difficulties that often seem to be overlooked by users when interpreting clusters or topics, and we developed ways to deal with these difficulties. In the case of interpreting topics, it is common practice to label them based on the top N words associated with each topic (Hecking & Leydesdorff, 2018). However, relying solely on these top N words presents challenges in distinguishing the meanings of topics, as common terms are likely to appear in the top N for multiple topics (Sievert & Shirley, 2014). As a result, interpreting topics based on high probability terms is insufficient. We learned that users need to consider both high-frequency terms and exclusive terms, that is, terms that occur frequently in a topic relative to their frequency of occurrence in other topics. This is consistent with the findings of Bishof and Airoldi (2012). The visualization tool that we used in our research simultaneously considers frequency and exclusivity, and therefore provides a valuable resource for interpreting topics.

Similarly, when it comes to interpreting clusters, relying simply on high-frequency terms can be problematic. These terms may sometimes mislead users in cluster interpretation. As a solution, combining information about journals and terms proved useful in interpreting clusters.

There are various issues that may be addressed in future research. In comparative analyses of different mapping approaches, it would be valuable to engage in more in-depth collaborative efforts with domain experts. Such collaborations could aim to provide insight into the usefulness of different approaches in specific research fields instead of trying to validate the outcomes of one specific approach. In light of the revealed strengths and weaknesses of TM and CC, another issue for future research could be to identify the underlying reasons for differences in the results provided by different methods and to find ways to strike a balance between the strengths and weaknesses of each method. Also, further research could study the temporal evolution of the structure of scientific fields, considering time-dependent information contained within words and citations.

**Acknowledgements**
We are grateful to Ismael Rafols, Alfredo Yegros and Pei Zhang for helpful discussions and suggestions, to Diane Gal and Karen Sipido for kindly helping us use the search strings for the delineation developed in Gal et al. (2015), to Qiao Zhao for kindly helping to scrutinize and explain topics, and to anonymous reviewers for valuable comments. We also appreciate the comments provided by the two anonymous reviewers, which have significantly contributed to improving this article.

**Author contributions**
Qianqian Xie: Conceptualization; Data curation; Formal Analysis; Investigation; Methodology; Software; Visualization; Writing – original draft.
Ludo Waltman: Conceptualization; Methodology; Supervision; Writing – review & editing.

**Competing interests**
The authors declare no competing interests.

**Funding information**
Qianqian Xie is financially supported by the China Scholarship Council.

**Data availability**
The data used in this study have been deposited in Zenodo: https://doi.org/10.5281/zenodo.8334551

**Code availability**
The code used in this study has been deposited in GitHub**:** https://github.com/Qianqian727/Preprocessing-procedure-and-LDA-traning.git.

# Appendix

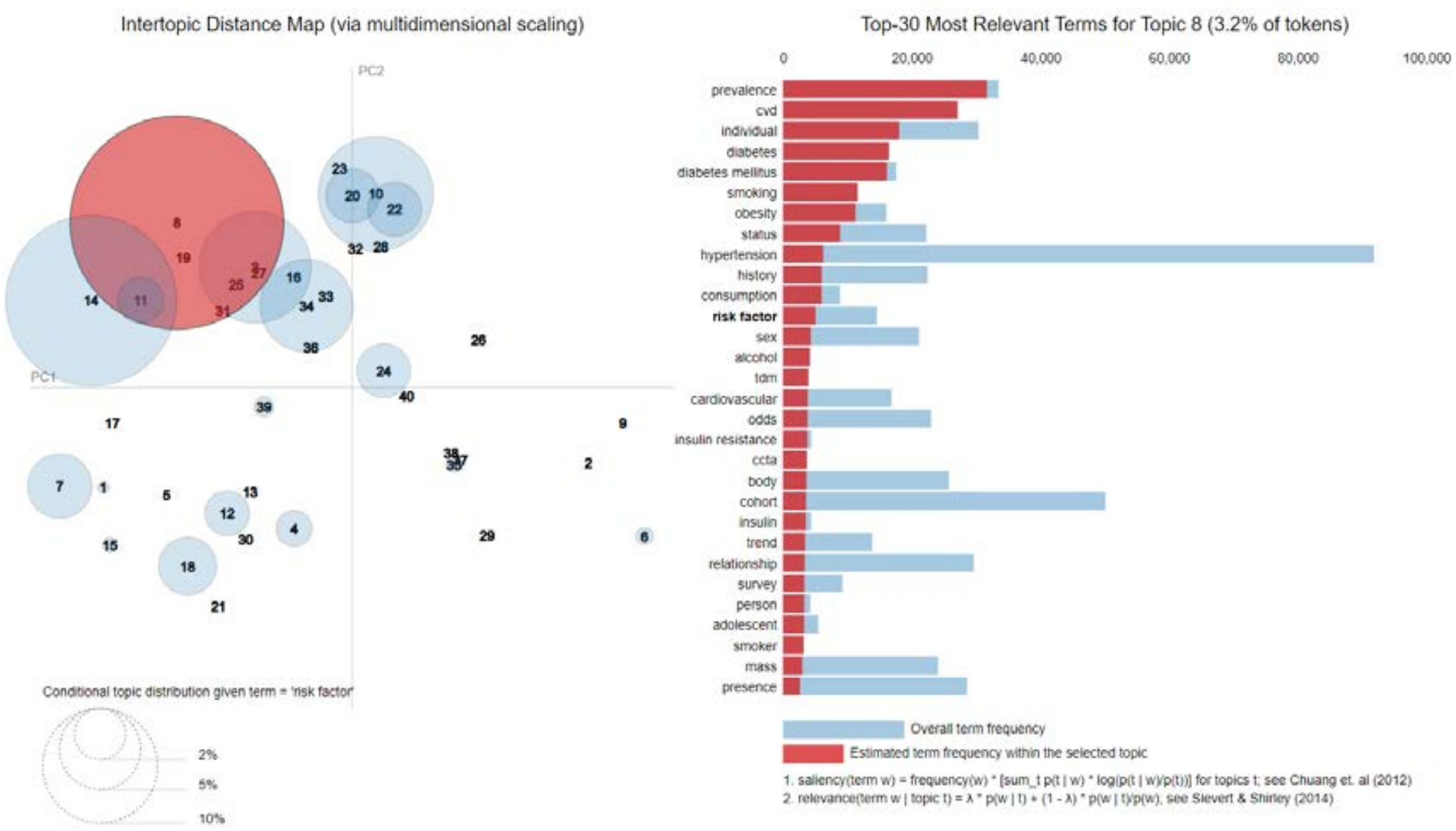

Figure A1. The distribution of risk factor term on the TM map